# A Dual EnKF for Estimating Water Level, Bottom Roughness, and Bathymetry in a 1-D Hydrodynamic Model

**Milad Hooshyar**
Department of civil, Environmental, and construction Engineering, University of Central Florida.
hooshyar.milad@knights.ucf.edu
**Stephen C. Medeiros**
Department of civil, Environmental, and construction Engineering, University of Central Florida.
**Dingbao Wang**
Department of civil, Environmental, and construction Engineering, University of Central Florida.
**Scott C. Hagen**
Department of civil and environmental engineering, Louisiana State University.

**Abstract**

Data assimilation has been applied to coastal hydrodynamic models to better estimate system states or parameters by incorporating observed data into the model. Kalman Filter (KF) is one of the most studied data assimilation methods whose application is limited to linear systems. For nonlinear systems such as hydrodynamic models a variation of the KF called Ensemble Kalman Filter (EnKF) is applied to update the system state in the context of Monte Carlo simulation. In this research, a dual EnKF approach is used to simultaneously estimate state (water surface elevation) and parameters (bottom roughness and bathymetry) of the shallow water models. The sensitivity of the filter to 1) the quantity and precision of the observations, and 2) the initial estimation of parameters is investigated in a 1-D shallow water problem located in the Gulf of Mexico. Results show that starting from an initial estimate of bottom roughness and bathymetry within a logical range and utilizing observations available at a limited number of gages the dual EnKF is able to improve the bottom roughness and bathymetry fields. The performance of the filter is sensitive to the precision of measured data, especially in the case of estimating Manning's $n$ and bathymetry simultaneously.

**Key words:** Parameter estimation; State estimation; Kalman Filter; Hydrodynamic model; Bathymetry estimation; Dual estimation

## 1. Introduction

Hydrodynamic models typically utilize numerical techniques; *i.e.,* finite element [*Hagen et al.*, 2000; *Tamura et al.*, 2014], finite difference [*Sadourny*, 1975], finite volume [*Mingham and Causon*, 1998; *Bradford and Sanders*, 2002], to solve a form of the Shallow Water Equations (SWEs). Given the initial water depths and velocities throughout the domain along with the boundary conditions, hydrodynamic models compute the deviation of the water surface and velocity for all predefined temporally and spatially discrete points. Hydrodynamic models, as with all numerical models, contain error due to both parametric and structural uncertainties [*Tatang et al.*, 1997]. Parametric uncertainty is primarily due to insufficient knowledge about bottom roughness [*Mayo et al.*, 2014], wind canopy [*Teeter et al.*, 2001], bathymetry [*Wilson et al.*, 2010; *Wilson and Özkan-Haller*, 2012], boundary and initial conditions, whereas structural uncertainty is mainly caused by poor representation of physical details, oftentimes due to insufficient model resolution [*Hagen et al.*, 2000; *Kim et al.*, 2014].

Methods to counteract these uncertainties begin with improvements to the characterization of: the geometric description of the domain through increased model resolution [*Hagen et al.*, 2000;

*Kim et al.*, 2014], and the parameters [*Schubert et al.*, 2008; *Medeiros and Hagen*, 2013]. When the best possible physical representation of the natural system is achieved, we are left with calibration and/or data assimilation to reduce model uncertainty.

In calibration, the set of model parameters are adjusted to maximize the agreement of model results and measured data [*Trucano et al.*, 2006]. In data assimilation, measured data are integrated into the numerical model as the simulation is performed in order to balance information from simulation and observation [*Wang and Cai*, 2008]. Filtering, a type of data assimilation, sequentially updates the state of the system based on available measurements. Among all available filtering methods, the Kalman Filter (KF) has been widely applied to deal with the uncertainty in numerical problems. The traditional KF updates the state of the system by minimizing the mean square error of the observed and measured state in linear systems. The KF is generalized for nonlinear systems by linearization of the state equation at each time step. This approach is called Extended Kalman Filter (EKF) [*Ljung*, 1979]. Although the EKF mitigates the linear model constraints of the traditional KF, it is not preferable for highly nonlinear systems due to the elimination of higher order moments of the error covariance equation [*Madsen and Cañizares*, 1999; *Evensen*, 2009].

Another approach for extending the KF to nonlinear systems is based on coupling KF and Monte Carlo simulation. This approach is called Ensemble Kalman Filter (EnKF) [*Evensen*, 2003]. Instead of one single state, an ensemble of states moves forward in time. The statistics of the state are calculated using the state ensemble with much less computational effort than with the EKF [*Evensen*, 2006]. The EnKF has previously been applied to water resources problems such as improving seasonal snowpack estimation [*Dechant and Moradkhani*, 2011], quantifying uncertainty of hydrologic forecasting [*Dechant and Moradkhani*, 2012], and estimating the state (*i.e.,* water surface elevation and/or velocity) or error of the hydrodynamic models [*Heemink and Kloosterhuis*, 1990; *Madsen and Cañizares*, 1999; *Babovic and Fuhrman*, 2002; *Sørensen et al.*, 2004; *Suga and Kawahara*, 2006; *Mancarella et al.*, 2008; *Tossavainen et al.*, 2008; *Ojima and Kawahara*, 2009; *Butler et al.*, 2012; *Karri et al.*, 2013; *Tamura et al.*, 2014] due to its simple conceptual formulation and relative ease of implementation [*Evensen*, 2003]. Particle Filter (PF) is an alternative to the EnKF for state and parameter estimation whose performance has been compared to the EnKF for hydrological forecasting by *Dechant and Moradkhani* [2012]. *Moradkhani et al.* [2012] improved the parameter search of the PF using Markov Chain Monte Carlo method which is also used by *Yan et al.* [2015] to improve soil moisture profile predictions.

Despite the extensive study of the EnKF for state estimation, there has been a relatively small number of studies for parameter estimation in hydrodynamic models. *Mayo et al.* [2014] used a Singular Evolutive Interpolated Kalman (SEIK) filter [*Pham*, 1996] to estimate Manning's *n* in the 2-D shallow water model ADCIRC. *Wilson et al.* [2010] used the EnKF to estimate bathymetry in the nearshore with the SHORECIRC hydrodynamic model. LiDAR-based elevation data contains errors when dealing with wet surfaces [*Hooshyar et al.*, 2015] and *Wilson and Özkan-Haller* [2012] presented an ensemble-based data assimilation approach to estimate river depth by utilizing depth-averaged velocity measurements. *Landon et al.* [2013] extended the work of *Wilson and Özkan-Haller* [2012] by using drifter-based velocity measurements to estimate bathymetry in the Kootenai River, Idaho using data assimilation.

Parameter estimation using KF-based approaches are mainly performed by 1) adding parameters to the state vector and treating them as the state variables (joint estimation or state augmentation) [*Wang et al.*, 2009; *Wilson et al.*, 2010; *Rafiee et al.*, 2011; *Wilson and Özkan-Haller*, 2012; *Landon et al.*, 2013] and 2) using dual filters in which state and parameter estimations are performed using two parallel filters (dual estimation) [*Wan and Nelson*, 1997; *Moradkhani et al.*, 2005]. The first approach is relatively easy to apply; however, it may produce numerical instabilities due to the high degrees of freedom as the number of unknowns increase [*Moradkhani et al.*, 2005].

In this paper a framework for parallel estimation of state (water level) and parameter (bathymetry and bottom roughness) is presented for 1-D SWEs. In fact, there are three filters working in parallel to estimate state and parameters, one for state (water level), one for Manning's n and another one for bathymetry. This filter is capable of individual or simultaneous estimation of bathymetry and bottom roughness. The Manning's n and bathymetry fields are estimated using function forms to reduce the computational burden of the filter. An extensive sensitivity analysis is performed to better understand the limitations of the method presented. This paper is organized as follows: Section 2 outlines the 1-D SWEs as the state transition model; Section 3 discusses the general framework of state and parameter estimation; and Section 4 describes numerical examples and sensitivity analysis.

## 2. Shallow Water Equations

In hydrodynamic models, when the horizontal length scale is much larger than the vertical, the SWEs can be utilized to solve for water levels and currents. The SWEs are typically a depth-integrated form of the Navier–Stokes equations. The set of the SWEs that includes the generalized wave continuity equation (GWCE) and the non-conservative momentum equation in one-dimension are given as follows [*Hagen et al.*, 2000]:

$$\frac{\partial^2 \eta}{\partial t^2} + \tau_0 \frac{\partial \eta}{\partial t} - g \frac{\partial}{\partial x}\left(h \frac{\partial \eta}{\partial x}\right) - (\tau - \tau_0)\frac{\partial}{\partial x}(uh) = 0 \tag{1}$$

$$\frac{\partial u}{\partial t} + g \frac{\partial \eta}{\partial x} + \tau u = 0 \tag{2}$$

where $\eta$ is the deviation of the free surface from the datum or initial water level, $u$ is the velocity, $\tau_0$ is a weighting parameter to balance the contribution of the primitive form of the SWEs and its GWCE form, $\tau$ is the shear stress due to the bottom friction and $h$ is the depth (i.e. bathymetry) that is referenced relative to a datum (e.g., NAVD88). It should be noted that the total water depth is expressed as $H = \eta + h$. The bottom roughness parameter, used to compute bottom stress, is typically applied in a quadratic, depth-dependent form [*Luettich et al.*, 1992]:

$$\tau/\rho_0 = C_f |u|u \tag{3}$$

where $C_f$ is the coefficient of friction, calculated based on Manning's *n* as in Equation 4:

$$c_f = \frac{gn^2}{\sqrt[3]{H}} \tag{4}$$

Using Equations 3 and 4, the bottom stress coefficient in the 1-D SWEs is calculated as:

$$\tau = \rho_0 \frac{gn^2}{\sqrt[3]{H}} |u|u \tag{5}$$

where $\rho_0$ is the reference water density and $g$ is the gravitational acceleration.

Solving the SWEs requires discretizing the domain and the PDEs presented in Equations 1 and 2 and iteratively solving them using a numerical method such as finite difference [*Smith*, 1985], finite element [*Zienkiewicz et al.*, 1977] or finite volume [*Versteeg and Malalasekera*, 2007]. In this study, the SWEs are solved using finite element method in space and finite difference in time. More detail regarding the numerical solution is given in section 4.2.

For even the simplest of real-world problems, given a perfect mathematical representation and infinitesimally small resolution, the solution of the SWEs would still contain some error regarding the uncertainty of the bottom roughness coefficient, bathymetry, boundary and initial conditions. Data assimilation provides a framework to incorporate measured data into the numerical model to mitigate the effects of these errors. If the data is robust and the errors are well-defined then data assimilation can enhance the process of modeling the physics of flow.

### 3. Methods

Kalman Filter [*Kalman*, 1960] is a data assimilation approach in which noisy measurements are utilized to better estimate the state of a linear system. The KF is a popular filter in the data assimilation field due to its simplicity and effectiveness. The KF and its variants generally consist of finding the model forecast in the current time step, incorporating observed data, and updating the model forecast based on the observations.

Extended Kalman Filter (EKF) is an extension of the KF to nonlinear systems whose application is restricted to slightly nonlinear systems due to its unstable results in highly nonlinear systems and its large computational requirements [*Miller et al.*, 1994; *Evensen*, 2003]. Ensemble Kalman Filter (EnKF) [*Evensen*, 2003] is another KF-based approach appropriate to deal with nonlinear systems. The EnKF is a Monte Carlo estimation of the KF in which the covariance matrix is represented by a large ensemble of states moving forward in time. In other words, instead of one single state vector at each time step, a large set of states (*i.e.,* state ensemble) are integrated in time and the moments of the probability density functions are estimated in each time step using the state ensemble. Below, we briefly review the basic ideas and algorithms of the KF and EnKF and present a special EnKF algorithm to simultaneously estimate state and parameters of the SWEs.

### 3.1. Kalman Filter (KF)

In a linear system, which is the main focus of the KF, the state ($s_t$) evolves throughout time following a linear function as in Equation 6. The observation ($d_t$) is also a linear function of state ($s_t$) as described in Equation 7. $A_t$ and $H_t$ are the state and observation transition models. $w_t$ and $v_t$ are zero-mean uncorrelated noise of the state and observation with the covariances of $\Sigma_s$ and $\Sigma_m$, respectively [*Arulampalam et al.*, 2002].

$$s_{t+1} = A_t s_t + w_t \tag{6}$$

$$d_t = H_t s_t + v_t \tag{7}$$

As describe in Fig. 1, the forecasted state in time $t$, $s_{t|t-1}$ (*i.e.,* the forecasted state in time step $t$ by utilizing all the observations so far and not including $d_t$), is determined using Equation 8. Subsequently, incorporation of the observation $d_t$ improves the estimate of state based on Equation 9. $K_t$, which is referred to as the Kalman gain, is a function of the state and observation covariances as described in Equation 10 and 11 [*Arulampalam et al.*, 2002].

$$s_{t|t-1} = A_t s_{t-1|t-1} \tag{8}$$

$$s_{t|t} = s_{t|t-1} + K_t \big[ d_t - H_t^\mathrm{T} s_{t|t-1} \big] \tag{9}$$

$$K_t = A_t \Sigma_{t|t-1} H_t \big[ H_t^\mathrm{T} \times \Sigma_{t|t-1} \times H_t + \Sigma_m \big]^{-1} \tag{10}$$

$$\Sigma_{t+1|t} = A_t \left[ \Sigma_{t|t-1} - \Sigma_{t|t-1} H_t \big[ H_t^\mathrm{T} \times \Sigma_{t|t-1} \times H_t + \Sigma_m \big]^{-1} \times H_t^\mathrm{T} \times \Sigma_{t|t-1} \right] A_t^\mathrm{T} + \Sigma_s \tag{11}$$

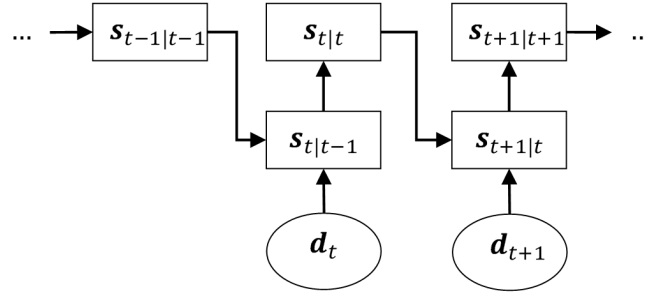

Fig. 1. Depiction of the Kalman Filter algorithm

The KF is developed upon the main assumptions that the system model is linear; however, when dealing with the SWEs, the system model is highly nonlinear (see Equation 1 and Equation 2). The KF has been extended for nonlinear systems by linearization of the system model (EKF). Its application however is limited to slightly nonlinear systems due to instabilities caused by discarding higher order moments of the error covariance equation in highly nonlinear systems [*Evensen*, 2009]. Another approach to extend the KF to nonlinear models is the EnKF which is discussed in the context of the SWEs model in the next section.

**3.2. Ensemble Kalman Filter (EnKF) for the SWEs**

To apply the EnKF to the SWEs, the problem must be first reformulated into a state-space framework. Based on the KF approach, the next state of the system should be a function of its current state (see Equation 6). However, in the SWEs, the next state (i.e. water surface elevation) is a function of both the current and the previous state due to the second derivative of water surface elevation with respect to time in the governing equations (Equation 1). When applying the EnKF and considering the water surface elevation in each time step $t$ as the state, the basis of the KF framework would be violated since the state would be a function of two previous time steps. In order to resolve this issue, the state of the system in each time step $t$, $\boldsymbol{s}_t$, is defined by concatenating the current and previous water surface elevation vectors ($\boldsymbol{\eta}_t$ and $\boldsymbol{\eta}_{t-1}$) as given in Equation 12.

$$\boldsymbol{s}_t = \{\boldsymbol{\eta}_t, \boldsymbol{\eta}_{t-1}\} \tag{12}$$

In state estimation using the EnKF, the error caused by parameterization and bathymetry is compensated for through adjustment of the water surface elevation and/or velocity. Using the EnKF, it is possible to not only enhance model performance by updating the state, but also computing more accurate parameters utilizing measured data. To do so, two main approaches are available; joint and dual estimation [*Moradkhani et al.*, 2005]. In joint estimation, the parameters are concatenated into the state vector and then treated as additional elements of the state vector. The state and parameters are estimated simultaneously using the KF (EnKF/EKF) in linear (nonlinear) systems. However, this approach is reported to have convergence problems [*Wan and Nelson*, 1997]. In addition, problems containing disparate types of parameters (i.e., bottom roughness and bathymetry in the SWEs) may result in numerical instabilities as the filter would treat all parameters in the same fashion. Instead of just a single filter, in dual estimation there are filters working in parallel, one for estimating the states and the other for parameters. Dual estimation was first introduced by *Wan and Nelson* [1997] to estimate both state and weights (i.e. parameters) of a Neural Network. In the following section, a dual EnKF for state and parameter estimation based on the approach suggested by *Moradkhani et al.* [2005] is presented for the SWEs where water surface elevation is defined as the state and the observations are also in form of water surface elevation. To demonstrate the method we assume there are M computational nodes throughout the domain, however the observations are available only in O ≪ M nodes and the ensemble size is denoted as N. This EnKF approach utilizes three filters, one for updating state (water surface elevation), one for bathymetry and one for Manning's *n* (bottom roughness).

In reality, the model's parameters (Manning's *n* and bathymetry) are spatially variable and each computational node holds a unique set of parameters. Modeling this spatial heterogeneity is computationally demanding since the parameters' ensembles would be significantly large (e.g., M × N matrices). The size of parameter ensemble can be considerably reduced by incorporating additional information to estimate the Manning's *n* and bathymetry field in a functional form. For instance, Manning's *n* can be approximated by a function $n_x = \mathcal{F}(x, \boldsymbol{\varphi})$, where $x$ is the location and $\boldsymbol{\varphi}$ is the set of hyperparameters of $\mathcal{F}$. Assuming $|\boldsymbol{\varphi}| = N_\varphi \ll M$, the dimension of the parameter ensemble of Manning's *n* reduces to $N_\varphi \times N$. Similarly, approximating bathymetry by $h_x = \mathcal{G}(x, \boldsymbol{\theta})$ will result in the contraction of the bathymetry ensemble size to $N_\theta \times N$, where $N_\theta \ll M$ denotes the dimension of the hyperparameter vector of $\mathcal{G}$. By doing so, the problem is simplified to estimation of the hyperparameters of the functional forms that represent the Manning's *n* or bathymetry.

The estimation at each time step $t$ is initiated by forecasting the hyperparameters of Manning's *n* and bathymetry functional form using a random walk as described in Equations 13 and 14.

$$\boldsymbol{\varphi}_{t|t-1} = \boldsymbol{\varphi}_{t-1|t-1} + \boldsymbol{q}_t \tag{13}$$
$$\boldsymbol{\theta}_{t|t-1} = \boldsymbol{\theta}_{t-1|t-1} + \boldsymbol{r}_t \tag{14}$$

where, $\boldsymbol{\varphi}_{t|t-1}$ ($\boldsymbol{\theta}_{t|t-1}$) is the forecasted ensemble of the hyperparameters in the Manning's *n* (bathymetry) functional form in time step $t$. $\boldsymbol{\varphi}_{t-1|t-1}$ ($\boldsymbol{\theta}_{t-1|t-1}$) denotes the updated ensembles in time step $t-1$. $\boldsymbol{q}_t$ and $\boldsymbol{r}_t$ are random noise with covariances $\sigma_n$ and $\sigma_h$, respectively. The Manning's *n* and bathymetry field (their value in each computational node) can be approximated by the predefined functions $\mathcal{F}$ and $\mathcal{G}$ using the forecasted hyperparameters (Equation 15 & 16). The forecasted ensemble of water surface ($\boldsymbol{\eta}_{t|t-1}$) is calculated by solving the continuity equation given in Equation 17;

$$\boldsymbol{n}_{t|t-1} = \mathcal{F}(\boldsymbol{\varphi}_{t|t-1}) \tag{15}$$
$$\boldsymbol{h}_{t|t-1} = \mathcal{G}(\boldsymbol{\theta}_{t|t-1}) \tag{16}$$
$$\boldsymbol{\eta}_{t|t-1} = \mathcal{C}(\boldsymbol{s}_{t-1|t-1}, \boldsymbol{u}_{t-1|t-1}, \boldsymbol{n}_{t|t-1}, \boldsymbol{h}_{t|t-1}) \tag{17}$$

where, $\mathcal{C}$ is the solution operator of general wave continuity equation (Equation 1), $\boldsymbol{u}_{t-1|t-1}$ is the ensemble of velocity field in time step $t-1$, $\boldsymbol{n}_{t|t-1}$ (Manning's coefficient) and $\boldsymbol{h}_{t|t-1}$ (bathymetry) are calculated in Equation 15 and 16. $\boldsymbol{s}_{t-1|t-1}$ is the ensemble of updated state in time step $t-1$. The forecasted ensemble of state is constructed by adding the forecasted water surface elevations for the current and previous time steps into one single vector based on Equation 18. The predictive observation ($\widehat{\boldsymbol{d}}_t$) is obtained by performing state and observation transition models to the forecasted state as shown in Equation 19.

$$\boldsymbol{s}_{t|t-1} = \{\boldsymbol{\eta}_{t|t-1}, \boldsymbol{\eta}_{t-1|t-1}\} \tag{18}$$
$$\widehat{\boldsymbol{d}}_t = \boldsymbol{H} \times \boldsymbol{s}_{t|t-1} \tag{19}$$

In this case, the operator $\boldsymbol{H}$ is an M × O binary matrix in which $H_{ij}$ is 1 if the $j^{th}$ element of observation vector corresponds to the state in computational node $i$. $\boldsymbol{H}$ is not necessarily a linear operator, however in the case of the SWEs and considering the water surface elevation as the state, it is a binary matrix since the observations are acquired explicitly in the form of water surface elevations. Having the Manning's *n*, bathymetry and predictive observation ensembles, the following forecasted covariance matrices are calculated;

$$\boldsymbol{\Sigma}_t^{d-d} = \overline{(\widehat{\boldsymbol{d}}_t - \overline{\widehat{\boldsymbol{d}}_t})(\widehat{\boldsymbol{d}}_t - \overline{\widehat{\boldsymbol{d}}_t})^{\mathrm{T}}} \tag{20}$$

$$\boldsymbol{\Sigma}_t^{s-d} = \overline{(\boldsymbol{s}_{t|t-1} - \overline{\boldsymbol{s}_{t|t-1}})(\widehat{\boldsymbol{d}}_t - \overline{\widehat{\boldsymbol{d}}_t})^{\mathrm{T}}} \tag{21}$$

$$\boldsymbol{\Sigma}_{t|t-1}^{\varphi-d} = \overline{(\boldsymbol{\varphi}_{t|t-1} - \overline{\boldsymbol{\varphi}_{t|t-1}})(\widehat{\boldsymbol{d}}_t - \overline{\widehat{\boldsymbol{d}}_t})^{\mathrm{T}}} \tag{22}$$

$$\boldsymbol{\Sigma}_t^{\theta-d} = \overline{(\boldsymbol{\theta}_{t|t-1} - \overline{\boldsymbol{\theta}_{t|t-1}})(\widehat{\boldsymbol{d}}_t - \overline{\widehat{\boldsymbol{d}}_t})^{\mathrm{T}}} \tag{23}$$

where, $\boldsymbol{\Sigma}_{t|t-1}^{d-d}$ is the error covariance matrix of the prediction and $\boldsymbol{\Sigma}_t^{s-d}$ denotes the cross covariance matrix of the state and prediction. $\boldsymbol{\Sigma}_t^{\varphi-d}$ represents cross covariance matrix of the prediction and Manning's *n* ensembles and. $\boldsymbol{\Sigma}_t^{\theta-d}$ is cross covariance matrix of the observation and bathymetry. $\overline{\widehat{\boldsymbol{d}}_t}$, $\overline{\boldsymbol{\varphi}_{t|t-1}}$ and $\overline{\boldsymbol{\theta}_{t|t-1}}$ denote the average of predictive observation, Manning's *n* and bathymetry ensembles. Based on the covariance values, the Kalman gain for updating the state ($\boldsymbol{K}$), Manning's *n* ($\boldsymbol{K}_n$) and bathymetry ($\boldsymbol{K}_h$) are calculated as;

$$\boldsymbol{K} = \boldsymbol{\Sigma}_t^{s-d}[\boldsymbol{\Sigma}_t^{d-d} + \boldsymbol{\Sigma}_m]^{-1} \tag{24}$$

$$\boldsymbol{K}_n = \boldsymbol{\Sigma}_t^{\varphi-d}[\boldsymbol{\Sigma}_t^{d-d} + \boldsymbol{\Sigma}_m]^{-1} \tag{25}$$

$$\boldsymbol{K}_h = \boldsymbol{\Sigma}_t^{\theta-d}[\boldsymbol{\Sigma}_t^{d-d} + \boldsymbol{\Sigma}_m]^{-1} \tag{26}$$

where, $\boldsymbol{\Sigma}_m$ is the covariance matrix of observation error constructed from the observation error variance $\sigma_m$. In order to update the state and parameters, it is essential to make an ensemble of observations. In most cases, the observations are only available in some specific computational nodes rather than for the entire domain. For the location in which the observation is available, the ensemble of observations ($\boldsymbol{d}_t$) is created by perturbation of the actual observation using a synthetic noise with the variance $\sigma_m$. Afterwards the state, Manning's *n* and bathymetry ensembles are updated based on Equations 27-29:

$$\boldsymbol{s}_{t|t} = \boldsymbol{s}_{t|t-1} + \boldsymbol{K}(\widehat{\boldsymbol{d}}_t - \boldsymbol{d}_t) \tag{27}$$

$$\boldsymbol{\varphi}_{t|t} = \boldsymbol{\varphi}_{t|t-1} + \boldsymbol{K}_n(\widehat{\boldsymbol{d}}_t - \boldsymbol{d}_t) \tag{28}$$

$$\boldsymbol{\theta}_{t|t} = \boldsymbol{\theta}_{t|t-1} + \boldsymbol{K}_h(\widehat{\boldsymbol{d}}_t - \boldsymbol{d}_t) \tag{29}$$

where, $\boldsymbol{s}_{t|t}$, $\boldsymbol{\varphi}_{t|t}$, and $\boldsymbol{\theta}_{t|t}$ are the updated ensembles of state, Manning's *n* and bathymetry, respectively. Since the updated state ensemble includes water surface in two subsequent time steps ($t$ and $t-1$), the state vector is detached to get water surface only in time step $t$. The depth-integrated velocity is calculated by solving the momentum equation (Equation 2);

$$\boldsymbol{u}_{t|t} = \mathcal{M}(\boldsymbol{u}_{t-1|t-1}, \boldsymbol{\eta}_{t-1|t-1}, \boldsymbol{n}_{t|t}) \tag{30}$$

where, $\mathcal{M}$ is the solution operator of the momentum equation (Equation 2), $\boldsymbol{u}_{t-1|t-1}$ is the ensemble of velocity in previous time step, $\boldsymbol{n}_{t|t} = \mathcal{F}(\boldsymbol{\varphi}_{t|t})$ is the updated ensemble of Manning's *n*.

Once again, the observed data are in the form of water surface elevations. It should be noted that depth-integrated velocity can also be added into the state vector; however, the velocity is updated separately since the momentum equation is solved using the previously updated water surface elevation. In this way, the velocity is indirectly affected by the observed data.

The algorithm presented here includes tuning parameters, namely the variance of observations error ($\sigma_m$), the variance of parameters error ($\sigma_n$, $\sigma_h$) and the ensemble size (N). *Mayo et al.* [2014] investigated the effects of observation error on Manning's *n* estimation by adding random noise (not exceeding 10% of magnitude). Their results show that the measurement error is an influential factor for parameter estimation. More importantly, the variance of Manning's *n* and bathymetry will affect the parameter estimation performance by altering the update rate of the parameters. The sensitivity of the dual EnKF to these parameters is discussed in the numerical examples presented in Section 4.

## 4. Numerical examples

In order to validate the performance of the presented dual EnKF approach, two types of numerical problems are considered. A linear 1-D state and parameter estimation problem is used to compare dual and joint estimation approaches (Case A). Additionally, a case study located in the Gulf of Mexico is considered in order to examine the performance of the dual EnKF for estimating Manning's *n* and bathymetry of a 1-D SWEs model. Based on this 1-D problem, four different cases (Cases B, C and D) are defined to assess the performance of the dual EnKF to estimate Manning's *n* and bathymetry both separately and simultaneously. Descriptions of all cases are listed in Table 1.

Table 1. The list of numerical examples

| Example | Description |
|---|---|
| **Case A** | Linear 1-D problem |
| **Case B** | known bathymetry and a single unknown Manning's *n* |
| B-1 | Similar to Case B, with different initial estimates of Manning's *n* |
| B-2 | Similar to Case B, with different values of observation error variance ($\sigma_m$) |
| B-3 | Similar to Case B, with different values of Manning's *n* error variance ($\sigma_n$) |
| B-4 | Similar to Case B, with different number of observation gages ($N_g$) |
| **Case C** | unknown bathymetry and a single known Manning's *n* |
| C-1 | Similar to Case C, with different initial estimates of Bathymetry |
| C-2 | Similar to Case C, with different values of observation error variance ($\sigma_m$) |
| C-3 | Similar to Case C, with different values of bathymetry error variance ($\sigma_h$) |
| **Case D** | unknown bathymetry and unknown Manning's *n* |
| D-1 | Similar to Case D, with different combination of Manning's *n* ($\sigma_n$) and bathymetry error variances ($\sigma_h$) |

### 4.1. Case A) Linear state-parameter estimation

As mentioned previously, the state and parameter estimation by the dual EnKF is performed in parallel whereas in the traditional approaches, both estimations are carried out using a single filter. To validate the concept of dual estimation, a simple 1-D differential equation is considered (Equation 31) that is linearized based on the finite difference method as represented in Equation 32.

$$\frac{dy}{dt} = H \times \cos 0.5t \tag{31}$$

$$y_{k+1} = y_k + \Delta t \times H \times \cos 0.5t \tag{32}$$

where, $y_k$ denotes the state in time step $k$, $H$ is the only parameter of the model and $\Delta t$ is the time resolution which is set to 0.005 s. In order to setup the parameter estimation problem, we assume $H_t$, the nominal value of H, is equal to 2 while the initial guess of $H$, denoted by $H_w$, is 1. Here, the goal is to estimate state $y$ and parameter $H$ using observations of state while $H$ is initially equal to $H_w$. The observations are synthetically generated by assuming perfect knowledge about the system and perturbation of the simulation results with a zero-mean normally-distributed error. The observation and parameter $H$ error variances are set to $1 \times 10^{-3}$ and $1 \times 10^{-6}$, respectively.

This problem is a linear state and parameter estimation that can be solved using the traditional KF-based joint estimation. As discussed earlier, joint estimation is performed by concatenating the parameter into the state vector. Alternatively, dual estimation is applied to estimate both state and parameter in parallel. Dual estimation utilizes two parallel filters, one for the state and another for the parameter, which are connected through the state transition function (Equation 32 in this case).

Fig. 2 shows the performance of dual and joint estimations both in terms of the state (Fig. 2a) and parameter (Fig. 2b). The joint estimation performed better for state estimation since the state quickly converges to the nominal value after only a small number of assimilations. For dual estimation, this happens after almost 30 assimilations. However, the dual filter performs better in parameter estimation, especially in the early stages of assimilation (compare the slope of dual and joint estimation curves in Fig. 2b). Although dual estimation converged faster in terms of parameter estimation, there are more fluctuations in the estimated parameter which are mainly due to the dynamic tradeoff between state and parameter estimation.

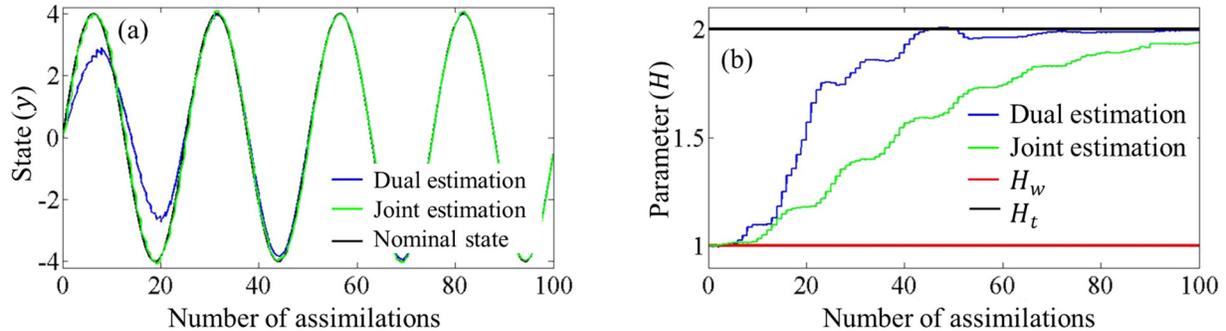

Fig. 2. The performance of dual and joint estimation in terms of state (a) and parameter (b) estimation for Case A.

### 4.2. SWEs cases

As mentioned in Table 1, Cases B, C and D are designed to examine the performance of the dual EnKF for estimating Manning's *n* and bathymetry of a 1-D SWEs model based on a transect located in the northern Gulf of Mexico (Fig. 3). The SWEs are solved numerically using the finite element method in space and finite difference discretization in time. The spatial resolution of the computational mesh is set to 5000 m (i.e., a uniform grid of points 5000 m apart) and the time resolution is 40 s. A 1-meter sinusoidal wave is imposed on the open Gulf boundary with a period of 1200 s. The near shore boundary is modeled as an absorbing boundary condition. The observations are synthetically generated by solving the SWEs for a nominal (i.e., true or actual) set of parameters (Manning's *n* and bathymetry) and are perturbed by zero-mean normally-distributed random error with variance $\sigma_m$. The simulations, both for data assimilation and observation generation, started from an initial state in which the water depth throughout the domain is equal to the mean sea level. The initial and boundary conditions are assumed to be known both in the assimilation procedure and observation generation in order to isolate their effects on the model results. In the base case, there are three simulated "gages" (orange triangles in Fig. 4) throughout the domain. The ensemble size in all cases is set to 30 and the state and parameters' ensembles are initiated by adding white noise with the corresponding variance to the initial values. For instance, the ensemble of Manning's *n* is initially generated by adding zero-mean normally-distributed error with a pre-defined variance ($\sigma_n$) to the initial guess of Manning's *n*.

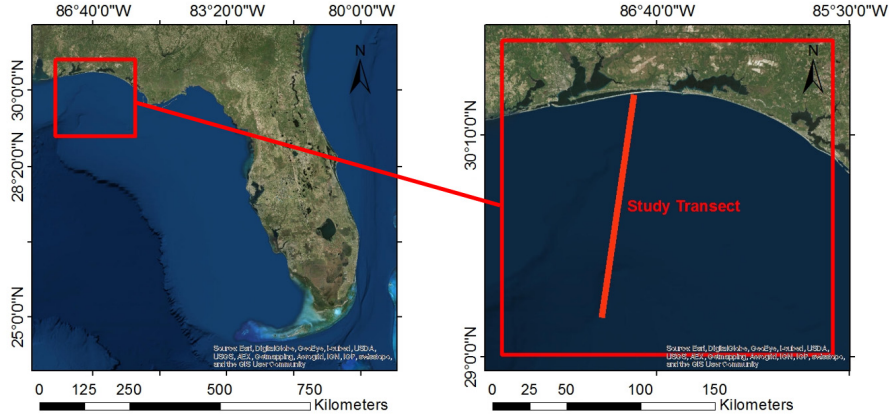

Fig. 3. The study area for Cases B, C and D along with the transect location

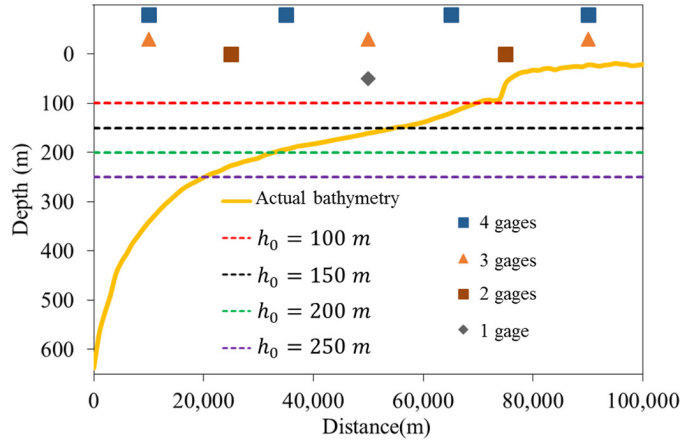

Fig. 4 – The nominal bathymetry (solid yellow line) and the initial guesses for bathymetry (dashed lines) used in Cases C and D. The gage locations are also shown for the experiments consisting of 1, 2, 3 and 4 gages. The gages are shown schematically to indicate their horizontal position.

**Case B) known bathymetry and a single unknown Manning's *n***

To demonstrate the performance of the dual EnKF for estimating Manning's *n*, a hypothetical case based on the 1-D transect in Fig. 3 is considered. For this case we assumed that the exact bathymetry is known as demonstrated in Fig. 4 and the Manning's *n* is unknown, although, it is assumed to be constant in space and time. The observation error variance is assumed to be $5 \times 10^{-4}$ and the Manning's *n* error variance is set to $5 \times 10^{-10}$. As discussed earlier, the Manning's *n* and bathymetry field are approximated as a functional form for computational proposes. Since in Case B, the Manning's *n* is spatially constant, a constant function is used to make the field of Manning's $n$. By doing so, the parameter ensemble has the dimension of $1 \times N$, where $N = 30$ is the ensemble size.

**Case B-1) Sensitivity to the initial guess of the Manning's $n$**

To perform parameter estimation using the dual EnKF algorithm, it is essential to have an initial guess of the parameters. A robust method should be capable of recovering the correct parameters starting from any initial guess within a logical range. To explore the performance of the parameter estimation, four experiments are performed. In each trial, one of the Manning's $n$ values in Table 2 is assumed to be the true representative of bottom roughness for the entire domain. Afterwards, starting from three other Manning's $n$, the nominal value is recovered. For instance, in Fig. 5b, the nominal Manning's $n$ is 0.01; each curve shows the estimated Manning's $n$ for three different initial guesses.

As shown in Fig. 5, the convergence time of the parameter estimation is not sensitive to the initial guess of Manning's $n$ as long as assimilation starts from a Manning's $n$ within a logical range. In fact, the convergence time is relatively constant. However, the initial guess did affect the magnitude of the update after each observation. When the difference between the initial guess and the actual value is relatively large, the magnitude of the updates are correspondingly large. This can lead to numerical instability for cases in which the initial guess is significantly different from the nominal Manning's $n$.

Table 2. Values of Manning's $n$

|  | Manning's $n$ |
|---|---|
| Class 1 | 0.005 |
| Class 2 | 0.01 |
| Class 3 | 0.015 |
| Class 4 | 0.02 |

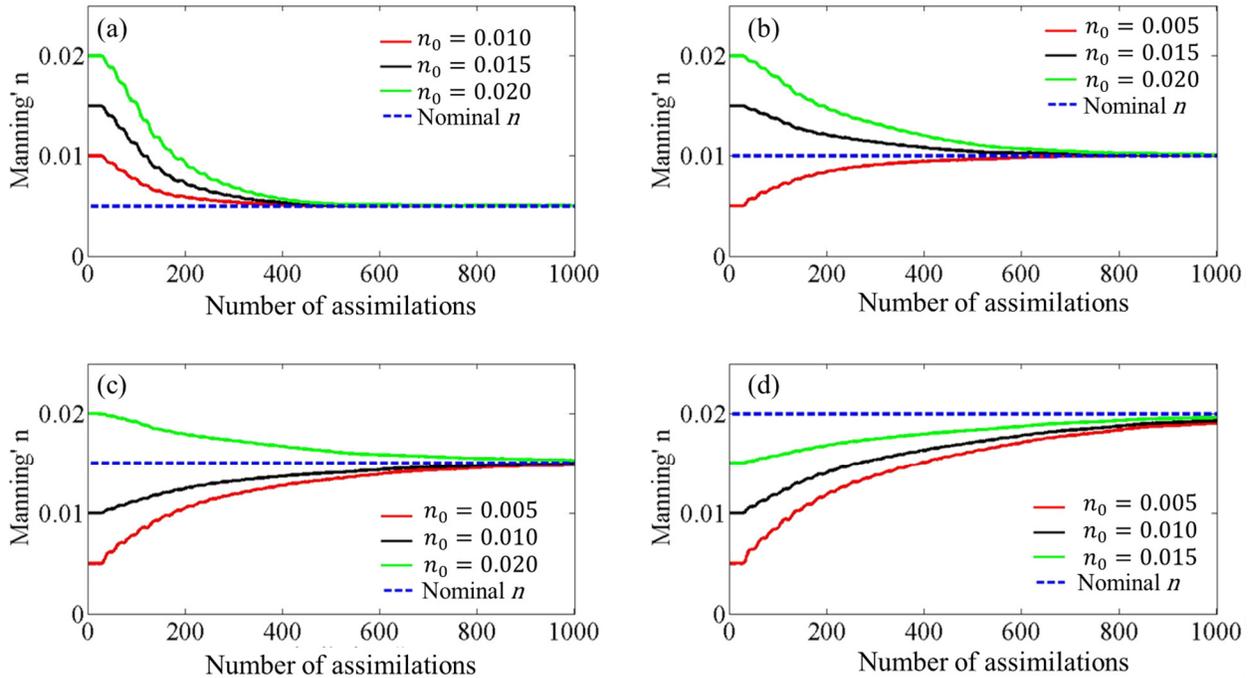

Fig. 5. The estimated Manning's *n* for Case B-1. The solid lines represent the estimated Manning's *n* for different initial guesses. Each subplot represents a different nominal Manning's coefficients which are shown by the dashed blue lines. The observation and Manning's *n* error variances are $5\times10^{-4}$ and $5\times10^{-10}$, respectively.

Since the water surface elevation represents the state of the model, the Mean Absolute Error (MAE) of the estimated and the nominal water surface approximates the accuracy of state estimation. Fig. 6 shows the performance of the dual EnKF in terms of state estimation based on the experiment presented in Fig. 5a. Since the initial condition of state is the same for all simulations, including the simulation for generating synthetic observations, the MAE is close to zero initially. However, large errors propagate into the model quickly after initiating the simulation. The dashed lines in Fig. 6 represent the MAE for the case when no assimilation is performed. Regardless of the initial guess, the error in state approaches to less than 0.05 m after approximately 500 assimilations, which is consistent with the convergence curve of Manning's *n* (Fig. 5a). The state estimation of the other experiments corresponding to Fig. 5b, c and d follow the same pattern and are not shown here.

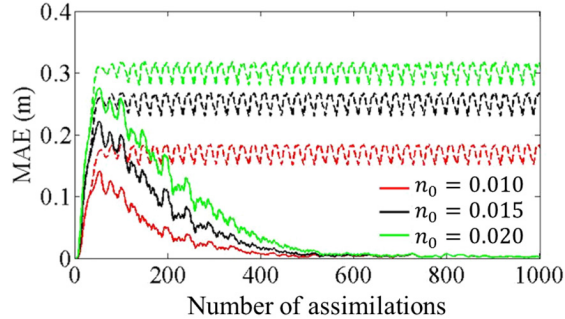

Fig. 6. The MAE of the estimated water surface elevation corresponding to Fig. 5a. The dashed lines represent the MAE without performing data assimilation. The solid lines show the MAE for different initial guesses of Manning's *n*. In all cases the nominal Manning's *n* is set to 0.005. The observation and Manning's *n* error variances are $5\times10^{-4}$ and $5\times10^{-10}$, respectively.

**Case B-2) Sensitivity to the observation error variance**

In this case, the sensitivity of data assimilation to observation precision is investigated. The precision of observation, represented by the observation error variance, is a key factor in data assimilation. Here, the observation error is represented as a zero-mean normally-distributed random number with variance $\sigma_m$. Smaller $\sigma_m$ represented higher observation precision.

Estimation of Manning's *n* results for different values of observation error variance ($5\times10^{-3}$, $5\times10^{-4}$, and $5\times10^{-5}$) are shown in Fig. 7a. Variance of observation error affects the convergence time; decreasing observation variance (*i.e.,* using more precise measurements) results in faster model convergence which is attributed to the increase of update size (i.e., the Kalman gain). When the observations are precise, their contribution to the update size is enhanced which leads to larger update magnitudes and faster convergence. The same pattern is observed in the state estimation. Fig. 8a shows the MAE of estimated water surface elevation in the experiment given in Fig. 7a. The precision of observation improves the accuracy of estimated state based on the same reasoning given for Manning's *n* estimation.

**Case B-3) Sensitivity to the Manning's *n* variance**

Besides the observation error variance, the variance of parameters' error (i.e., Manning's $n$) also plays an important role in the dual EnKF. These two variances determine the magnitude of update size after each observation. Fig. 7b shows the results of Manning's $n$ estimation using three different Manning's $n$ variances ($5\times10^{-10}$, $5\times10^{-11}$, and $5\times10^{-12}$). Generally, the response of the data assimilation to Manning's $n$ error variance is inversely related to that of the observation error variance. Larger Manning's $n$ error variance results in faster convergence due to larger update size. However, relatively large variances can result in instabilities when the magnitude of the update in each assimilation gets larger. Fig. 8b shows the MAE of estimated water surface elevation for different Manning's $n$ error variances based on the cases presented in Fig. 7b. Similar to the estimated Manning's $n$ (Fig. 7b), the Manning's $n$ error variance is directly related to the accuracy of state estimation.

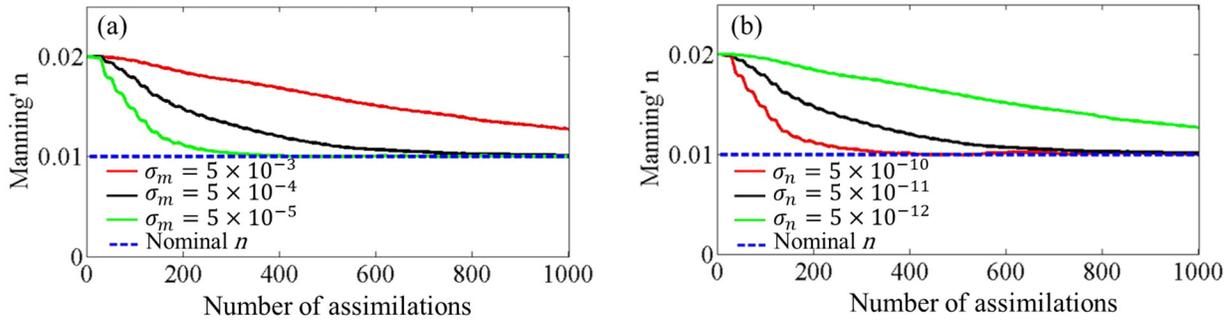

Fig. 7. The effect of observation and Manning's $n$ error variances on the estimated Manning's $n$ in Cases B-2 and B-3. The dashed blue lines represent the nominal Manning's $n$ and the solid lines are the estimated ones for different values of observation (a) and Manning's $n$ (b) error variances. In both cases the nominal and initial guess of Manning's $n$ are set to 0.01 and 0.02, respectively.

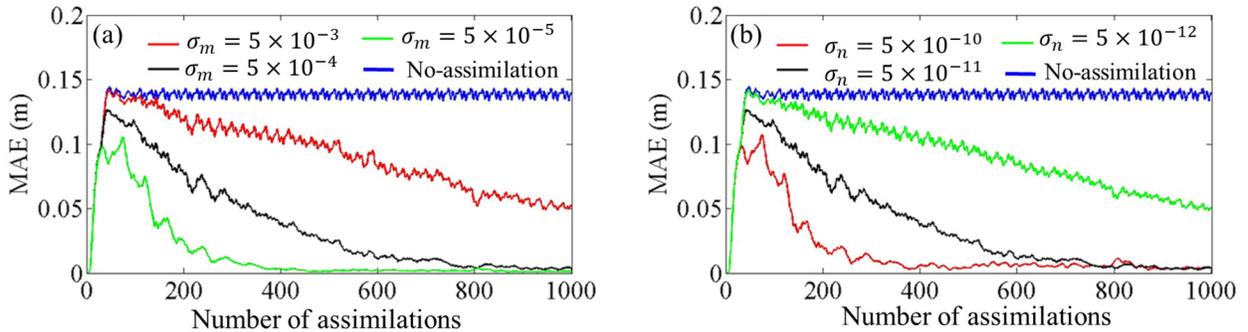

Fig. 8- The effect of observation (a) and Manning's $n$ (b) error variances on the MAE of the estimated water surface elevation for Cases B-2 and B-3. The blue lines represent the MAE without performing data assimilation. In both cases the nominal and initial guess of Manning's $n$ are set to 0.01 and 0.02, respectively.

**Case B-4) Sensitivity to the number of observation gages**

The amount of available measured data (*i.e.,* number of gages) is also a crucial factor in the EnKF. In Fig. 9, the assimilation results for 1, 2, 3, and 4 gages (the location of gages are shown

Fig. 4) is presented both in terms of the estimated Manning's *n* (Fig. 9a) and state (Fig. 9b). In general, comparing Fig. 9 with Fig.s 7 and 8 indicates that the data assimilation is more sensitive to the precision of observed data rather than its quantity.

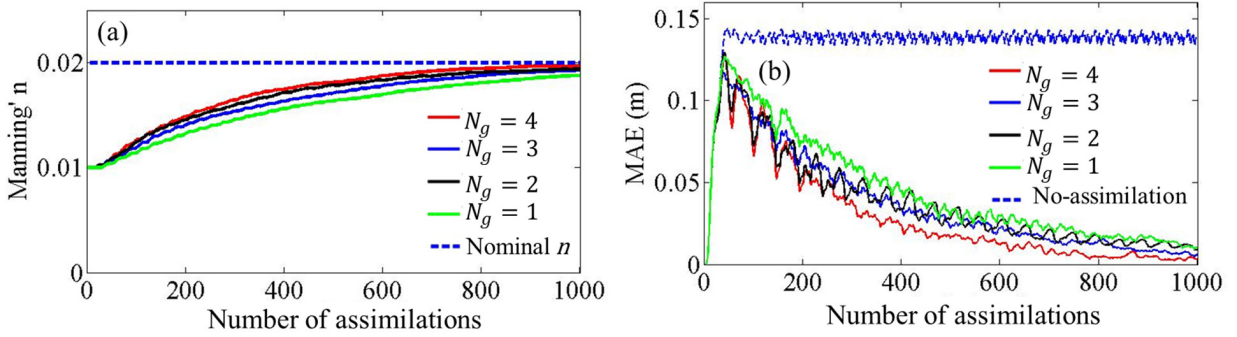

Fig. 9. (a) The estimated Manning's *n* in Case B-4 for different number of gages. The dashed blue line is the nominal Manning's *n*. (b) The MAE of the water surface elevation for different number of gages. The dashed blue line represents the MAE without performing data assimilation. The initial guess and nominal Manning's *n* are set to 0.01, 0.02. The observation and Manning's *n* error variances are $5\times10^{-4}$ and $5\times10^{-10}$, respectively.

**Case C) unknown bathymetry and a single known Manning's *n***

In Case C, the algorithm presented in Section 3.2 is applied to estimate bathymetry in a 1-D problem with given Manning's *n*. For computational purposes, the bathymetry field is assumed to be a piecewise linear function along the domain with 4 transition points located at $x = 0, 10, 45, 76 \ km$ from the Gulf boundary. Then, the problem of bathymetry estimation is simplified to finding the elevation in the transition points (i.e., the hyperparameters of bathymetry functional form), since the elevation of all computational nodes would be a linear function of depth in adjacent transition points (see Fig. 10 and Equation 33). By doing so, The size of the bathymetry ensemble reduces to $4 \times $ N, where N = 30 denotes the ensemble size. After each assimilation the hyperparameters of the bathymetry functional form (i.e., $\boldsymbol{\theta}_{t|t}$ in Equation 29 which represents the elevation in transition points) are updated and the elevation of any computational node $c$, located between transition points $i$ and $i + 1$, is calculated as;

$$h^c_{t|t} = \theta^i_{t|t} + \left(\theta^{i+1}_{t|t} - \theta^i_{t|t} \big/ l_{i,i+1}\right) l_{i,c} \quad \forall c | \ x_i < x_c \leq x_{i+1} \tag{33}$$

where, $\theta^i_{t|t}$ is the updated elevation of $i^{\text{th}}$ transition point (see Equation 29) and $l$ denotes the horizontal distance between points which are schematically illustrated in Fig. 10. $x$ denotes the horizontal distance of points from the open Gulf boundary. Increasing the number of transition points would result in more accurate bathymetry, however, it is more computationally expensive.

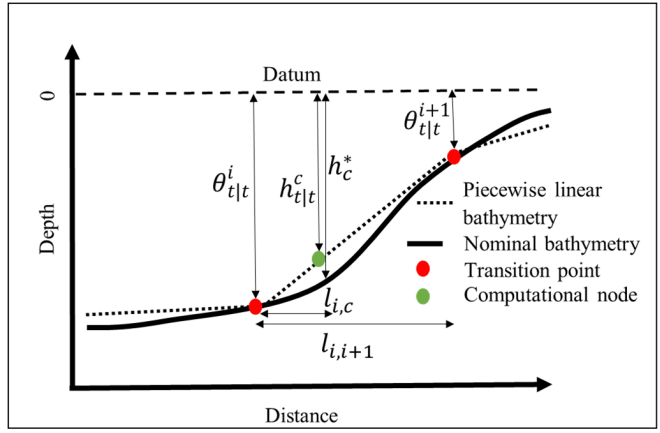

Fig. 10. Schematic of piecewise linear function for bathymetry. Solid black line shows the nominal bathymetry and the dashed line shows the estimated bathymetry approximated as a linear function of elevation in two adjacent transition points (red dots). Green dot is a computational node whose estimated elevation ($h^c_{t|t}$) is a linear function of elevation in $i$ and $i+1$ ($\theta^i_{t|t}$ and $\theta^{i+1}_{t|t}$) as shown in Equation 33. $h^*_c$ is the nominal elevation of node $c$ and $l$ denotes the horizontal distance between points.

For this experimental case, since the nominal bathymetry is given, the performance of estimated bathymetry in time $t$ is measured by averaging the ratio of elevation error at all computational nodes (Equation 34);

$$P_t = \frac{\sum_{c=1}^{M} \left| \left(h^*_c - h^c_{t|t}\right) / h^*_c \right|}{M} \tag{34}$$

where, M is the number of computational nodes, $h^*_c$ is the nominal bathymetry in computational node $c$ and $h^c_{t|t}$ is the estimated bathymetry for that specific node which corresponds to the updated hyperparameters in time $t$ ($\boldsymbol{\theta}_{t|t}$).

**Case C-1) Sensitivity to the initial guess of bathymetry**

Fig. 11 shows the performance of bathymetry estimation using four different initial depths, including constant depth of 100, 150, 200 and 250 m as shown in Fig. 4. The variance of bathymetry and observation error are set to 0.025 and $5\times10^{-4}$, respectively. In this case, there are three observation gages throughout the region as shown in Fig. 4. Using the dual EnKF, even starting from a poor initial guess, it is possible to recover the nominal bathymetry. For instance, starting from constant depth of 150 m, which corresponded to an average bathymetry error of 0.75 (i.e., on average the bathymetry have 75% error in each computational point), the bathymetry is estimated to an accuracy of almost 10% (Fig. 11a).

The performance of the dual EnKF in terms of state estimation is shown in Fig. 11b for two different initial guesses of bathymetry (100 m and 200 m), using the MAE of estimated water surface elevation as a proxy for state accuracy. After approximately 600 assimilations, the MAE reduced to less than 0.1 m.

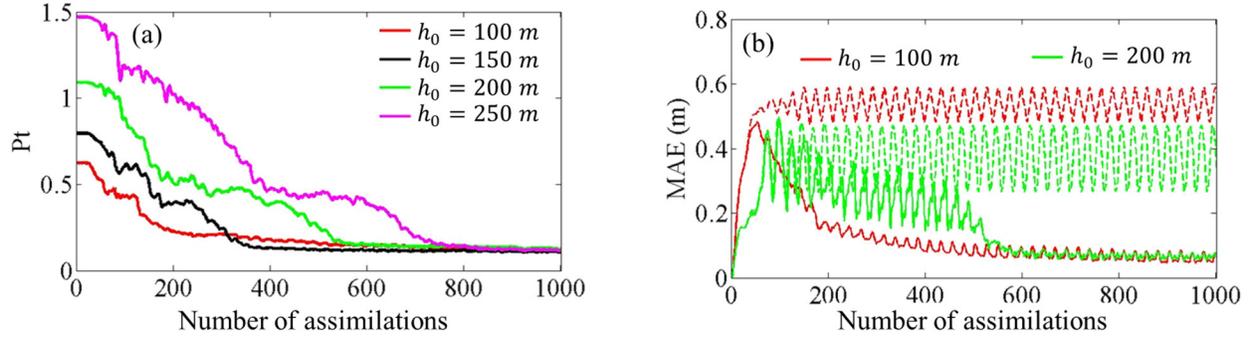

Fig. 11. (a) The error of estimated bathymetry for different initial guesses of bathymetry in Case C. The initial bathymetry is assumed to be flat with depth $h_0$ throughout the domain. $P_t$ is an indicator of the error in the estimated bathymetry and is given by Equation (34). (b) The MAE of water surface elevation (accuracy of state estimation) for two selected initial guesses of bathymetry (100 m and 200 m). The dashed lines represent the MAE without performing data assimilation and the solid lines show the MAE for different initial guesses. The observation and bathymetry error variances are $5\times10^{-4}$ and 0.025, respectively.

**Cases C-2 & C-3) Sensitivity to the observation and bathymetry error variances**

Sensitivity to bathymetry and observation error variances is investigated by setting observation error variance to $5\times10^{-3}$, $5\times10^{-4}$, and $5\times10^{-5}$ and bathymetry error variance to $1.5\times10^{-2}$, $2.5\times10^{-2}$, and $3.5\times10^{-2}$. The results followed the same trend as shown in Case B. The performance of the dual EnKF for these different variances is shown in Fig. 12 (bathymetry estimation accuracy) and Fig. 13 (state estimation accuracy).

Using an observation error variance of $5\times10^{-3}$, the dual EnKF does not converge both in terms of bathymetry (Fig. 12a) and state (Fig. 13a). The same behavior regarding bathymetry and state estimation is observed when setting the bathymetry error variance to $1.5\times10^{-2}$ (Fig. 12b and 13b).

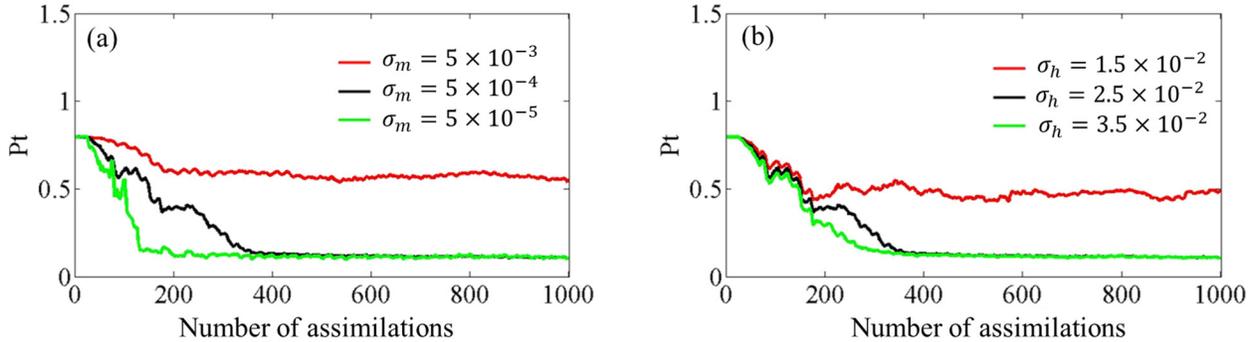

Fig. 12. The effect of observation and bathymetry error variances on the estimated bathymetry in Cases C-2 and C-3. (a) The error of estimated bathymetry for various observation error variances ($\sigma_m$). (b) The impact of varying bathymetry error variance ($\sigma_h$) on the error of bathymetry estimation. $P_t$ is an indicator of the error in the estimated bathymetry and is given by Equation (34). In all cases the initial bathymetry is flat with a depth of 150 m.

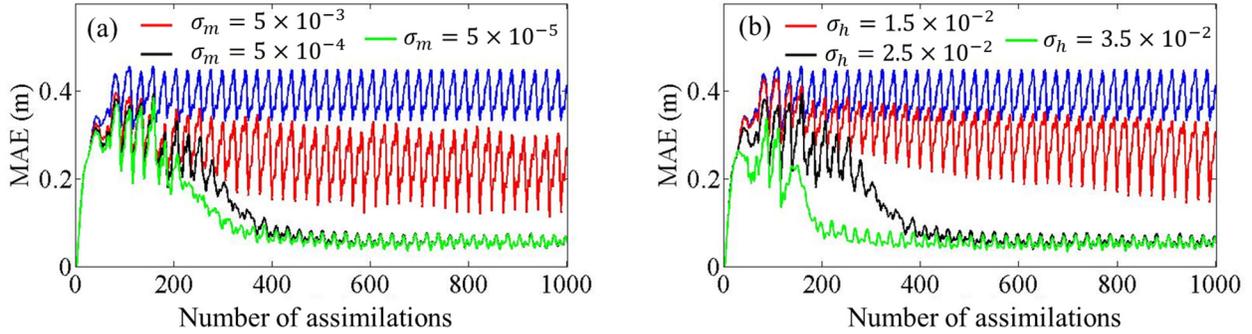

Fig. 13. The effect of observation (a) and bathymetry (b) error variance on the MAE of the estimated water surface elevation corresponding to experiments presented in Fig. 12. The blue lines represent the MAE without performing data assimilation. In all cases the initial bathymetry is flat with the depth of 150 m.

**Case D) unknown bathymetry and unknown Manning's *n***

When dealing with estimation of multiple parameters, the main challenge is to set up the parameter variances in order to have balanced parameter estimation. Improper variances could result in biased parameter estimation in which the model compensates for error caused by one parameter by changing another.

The effects of bathymetry and Manning's *n* variances are investigated for a problem in which both bathymetry and Manning's *n* are unknown. In this case, the setup for the Manning's *n* field is similar to Case B, the initial guess for Manning's *n* is assumed to be 0.02, whereas the nominal value is 0.01. Similar to Case C, a piecewise linear function with 4 transition points is utilized to approximate the bathymetry. The initial guess for bathymetry is a constant depth of 100 m along the domain. Nine combinations of Manning's *n* and bathymetry error variances are tested to investigate to accuracy of estimated parameters. The performance regarding bathymetry is measured by averaging the ratio of elevation error in all computational nodes (Equation 34).

As shown in Fig. 14, as $\sigma_n$ decreases, the convergence speed of Manning's *n* reduces. This trend continues to the point where for larger values of $\sigma_h$ the model becomes unstable (Fig. 14c). In terms of estimated state (Fig. 15), the dual EnKF has robust performance; in most cases the water surface elevation error reduces to 0.05 m after approximately 700 assimilations. An interesting observation is that the dual EnKF has robust performance in terms of state estimation even when the parameters are estimated poorly (Compare Fig. 14c and Fig. 15c). This is attributed to the combined effects of parameters on the state. The error in water surface elevation (state) caused by Manning's *n* is likely to be compensated for by an unrealistic bathymetry and vice versa. This observation emphasizes the necessity of selecting proper variances for parameters to avoid achieving good results in terms of state by utilizing an unrealistic set of parameters.

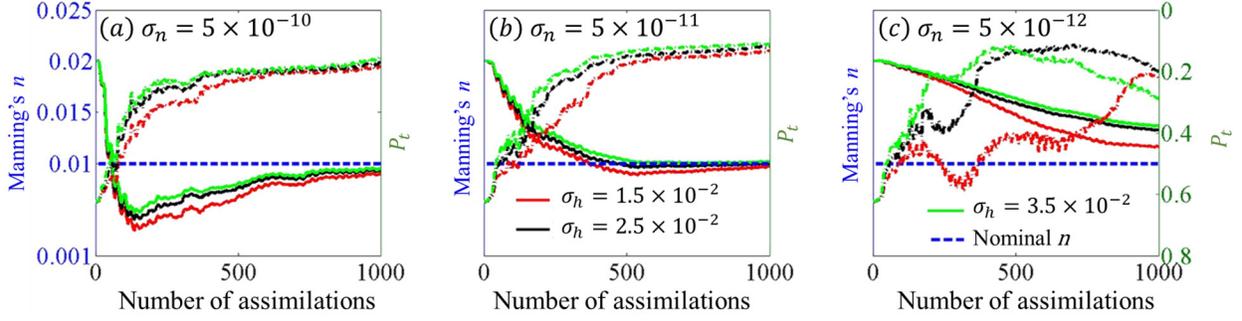

Fig. 14. Bathymetry and Manning's *n* estimation for Case D with varying Manning's *n* and bathymetry variances. Dashed lines show the error of the estimated bathymetry ($P_t$ from Equation 34) which can be read on the right vertical axis. The solid lines represent the estimated Manning's *n*. Dashed blue lines show the nominal Manning's *n* in each case. In each subplot, the Manning's *n* variance is constant ($5\times10^{-3}$ in a, $5\times10^{-4}$ in b and $5\times10^{-5}$ in c). The corresponding bathymetry error variance for each line color is given in the legend. The closeness of the solid lines to the blue dashed line (nominal Manning's *n*) indicates better performance in Manning's *n* estimation. Smaller values of dashed lines (closer to zero) indicates better bathymetry estimation. The observation error variance is set to $5\times10^{-4}$

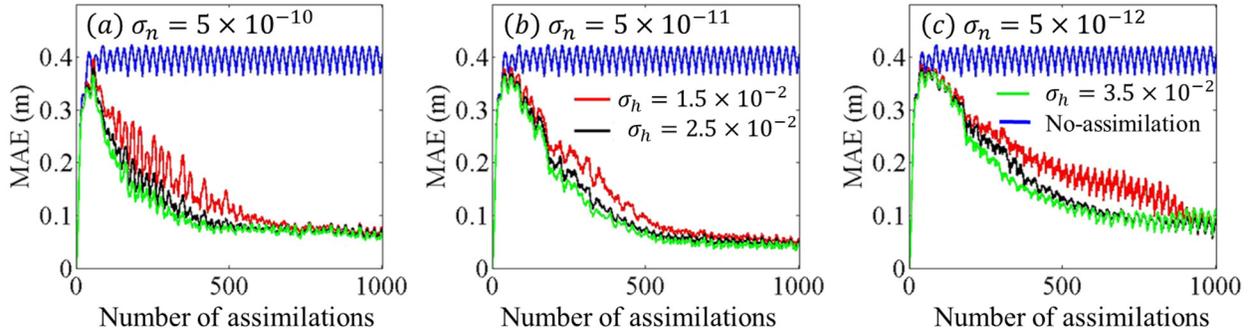

Fig. 15. The MAE of the estimated water surface elevation for Case D with varying Manning's *n* and bathymetry variances corresponding to Fig. 14. The blue lines represent the MAE without performing data assimilation. In each subplot, the Manning's *n* variance is constant ($5\times10^{-3}$ in a, $5\times10^{-4}$ in b and $5\times10^{-5}$ in c). The corresponding bathymetry error variance for each line color is given in the legend. The observation error variances is set to $5\times10^{-4}$

## 5. Conclusion

The handling of parameter uncertainty in hydrodynamic models is crucial in coastal modeling due to its effect on the accuracy of model results. Calibration and data assimilation are two approaches to address these errors. Using data assimilation, it is possible to not only enhance model results but also to optimally estimate model parameters (i.e. bottom roughness, bathymetry, etc.). In this paper, a dual EnKF is applied to the 1-D shallow water equations (SWEs) to estimate water surface elevation, bathymetry and bottom roughness. The method includes multiple filters working in parallel, each one responsible for a single type of parameter and state.

The results indicate that the initial guess of parameters has relatively minor effects on the convergence of the filter; however, the filter is highly sensitive to both observation and parameter error variances. For instance, starting from a constant depth across the domain it is possible to estimate bathymetric depths close to the nominal values. Observation error variance (an indicator of observation precision) has negative effects on convergence. The response to the parameters' variance is more complicated. Estimating just a single parameter (bathymetry or Manning's *n*), the parameter variance haspositive effect on convergence. However, high variance causes instabilities since the magnitude of update gets relatively large. For estimating both bathymetry and Manning's *n*, improper parameter variances can result in biased estimation, meaning that the filter tries to compensate for the error caused by one parameter by changing the other. The error variance of parameters should reflect the level of uncertainty in the initial guess. In other words, when using a rough initial guess, one needs to apply a relatively high variance to be able to find the correct parameter through data assimilation. When dealing with a problem with multiple parameters, it is challenging to appropriately set the variances in order to achieve a balanced update. The presented work can be used as a guide to select variances for parameter estimation in hydrodynamic models as their effect is investigated in a variety of estimation cases. However, the characteristics of the user's specific problem in terms of the accuracy of initial guesses should also be taken into account when selecting proper error variances.

The proposed method is shown to reduce uncertainty of bathymetry and bottom roughness by utilizing the measured water surface elevation. This improves the performance of hydrodynamic models in sensitive regions such as marshlands and nearshore area that typically have high levels of uncertainty in measured bottom roughness and/or bathymetry. Although the method is applied to a 1-D SWEs model, it is possible to apply the dual filter approach to 2-D models and future work will address this task. The main challenge with respect to the application of the dual EnKF in 2-D problems is the amount of available observation in real problems. The dual EnKF requires relatively more observed data than the traditional EnKF since state, Manning's n and bathymetry are being updated simultaneously. However the filter can utilize even limited amount of observations and proportionally improve the parameters and state estimation. In that case, although we won't achieve the exact parameters and perfect state, we will get at least better estimations. In addition, in real-world problems, the uncertainty of parameters is usually limited both spatially and in range. For instance, the Manning's n in marshlands, which often covers a limited portion of the spatial domain, is usually subjected to high uncertainty, however the upper and lower bounds of Manning's n in those regions are given with acceptable accuracy. In other words, we do not need to estimate the parameters in the whole domain which requires extensive observations. Instead, we can utilize the limited data to focus only on regions with high uncertainty.

**Acknowledgement**

This research was funded in part under Award NA10NOS4780146 from the National Oceanic and Atmospheric Administration (NOAA) Center for Sponsored Coastal Ocean Research (CSCOR), the NASA Kennedy Space Center, Ecological Program, Climate Adaptation Science Investigators (CASI) project (Award number: IHA-SA-13-006), and the Louisiana Sea Grant Laborde Chair endowment. The statements and conclusions are those of the authors and do not necessary reflect the views of NOAA, NASA, Louisiana Sea Grant, or their affiliates.